\documentclass[10pt, a4paper, twocolumn]{IEEEtran}
\usepackage{amsmath,epsfig,amssymb,verbatim,amsopn,cite,subfigure,multirow, mathrsfs}
\usepackage{balance}

\usepackage{gensymb}
\allowdisplaybreaks

\usepackage[usenames,dvipsnames]{color}
\usepackage[all]{xy}  %%% used to make block diagram
\usepackage{url}
\usepackage{amsfonts}
\usepackage{amssymb}
\usepackage{epsfig}
\usepackage{amsmath,epsfig,amssymb,verbatim,amsopn,cite,subfigure,multirow}

\usepackage{amsmath,epsfig,amssymb,verbatim,amsopn,cite,subfigure,multirow}
\usepackage{balance}
\usepackage[usenames,dvipsnames]{color}
\usepackage[all]{xy}  %%% used to make block diagram
\usepackage{url}
\usepackage{amsfonts}
\usepackage{amssymb}
\usepackage{epsfig}
\usepackage{cite}
\usepackage{graphicx}
\usepackage{epstopdf}
\usepackage{enumerate}

\usepackage[utf8]{inputenc} 
\usepackage[T1]{fontenc}

\title{Barometers Can Hear, and Sense Finger Taps}

\author{Alireza Hafez, Dorsa Nahid, and Majid Khabbazian
  \thanks{%
    A. Hafez, D. Nahid, and M. Khabbazian are with the Department of Electrical and Computer Engineering, 
    University of Alberta, Edmonton, Canada
    (Email: \{hafez, dorsa, mkhabbazian\}@ualberta.ca).}
  }

\begin{document}
\maketitle
\begin{abstract}
    Most modern smartphones are equipped with a barometer to sample air pressure.
    Accessing these samples is deemed harmless, hence does not require any permission.
    In this work, we show, however, that these samples can reveal sensitive information in smartphones with ingress protection.
    For the first time, it is shown that barometer samples, even at a low rate of 25 Hz, 
    can leak information about the smartphone’s speaker activity.     
    Specifically, we use these samples to detect with high accuracy ($\geq 95\%$)
    whether the smartphone's speaker is silent or playing a sound such as a ringtone.
    In addition, we use the samples to detect the activity of an external speaker.
    Finally, we show that low-rate barometer samples can be used to 
    1)~detect touchscreen finger taps with $100\%$ accuracy, and 
    2) gain information about the positions of finger taps.

  \end{abstract}

\begin{IEEEkeywords}
Smartphone, barometer, side-channel inference.
\end{IEEEkeywords}

\section{Introduction}

    Smartphones have become an integral part of our daily lives.
    We have been increasingly relying on them for various tasks including communications, 
    navigation, fitness tracking, gaming, social networking and banking.
    To support this wide set of tasks, smartphones have been equipped with many sensors
    ranging from microphones and cameras to GPS, accelerometers, gyroscopes and barometers.

    The abundance and sensitivity of data collected by sensors in smartphones is a major privacy concern.
    To address this concern, smartphones have put in place permission mechanisms that allow users to control apps' access to 
    many of the these sensors.
    In Android, for example, apps must get permissions from users either at install time or at run time 
    to access sensitive sensors.
    In addition, access to sensitive sensors may be taken away when the app goes to the background.
    For instance, an app in the background does not get access to the touchscreen input, 
    as the input may contain sensitive information such as passwords or texts that belong only to the foreground app.
    
    There are sensors whose access requires no permission or user notification, because their data is deemed to be safe.
    One such sensor is barometer, which is now found in most mid-range to high-range smartphones.
    In addition to measuring air pressure, barometers can be used as altimeters since pressure data can be translated to altitude.
    Sensitivity of barometer samples makes it possible to detect altitude changes as little as one meter.
    Because of this sensitivity, many apps including floor level detection in indoor tracking rely on barometer sensors.
    Barometers can also be used in low-power outdoor tracking as they consume significantly lower power than GPS.

    In this work, we investigate whether barometer can be used as a side channel to gain sensitive information. 
    A major challenge in doing so is the low sampling rate of smartphone's barometers, currently limited to 25Hz~\cite{sankaran2014using}.
    Despite this limitation, we show two possible side channel information leaks in smartphones with ingress protection. 
    First, we show for the first time that barometer samples reveal information about the smartphone's speaker activity, as well as
    the activity of an external speaker.
    Second, we show that the samples leak information about the occurrence  of 
    a finger tap (a gentle touch on the screen with a finger), and their positions.

      \textbf{Paper Organization.}
       The remainder of this paper is organized as follows.     
       A review of related work is presented in Section~\ref{sec:Rel}.       
       A background is provided in Section~\ref{sec:back}. % and discuss the threat model in~\ref{sec:Thr}.
       We discuss our initial observations in Section~\ref{sec:Obs}, 
       and present our experimental methodology and results in Section~\ref{sec:Exp}. 
       We discuss some defense solutions in Section~\ref{sec:Def}.
       Finally, we conclude and present future research directions in Section~\ref{sec:Con}.

\section{Related work}
\label{sec:Rel}
  
  Barometric air pressure measurements by smartphones' barometers can be used to estimate altitude, 
  and detect altitude changes of as little as one meter~\cite{SankaranZGACP14}.
  Because of this, barometers on smartphones have been used in many apps
  such as aiding GPS~\cite{ZhangEZCL12}, detecting floor level for indoor positioning~\cite{VaniniG13}, 
  and improving calorie estimation in fitness apps~\cite{SankaranZGACP14}.

  Smartphone barometers have not been explicitly used in the literature for side channel attacks.  
  However, existing work suggest that barometer's measurements leak information about users' activities and surroundings. 
  Sankaran et al. ~\cite{SankaranZGACP14} showed that barometer data can be used to detect user's activities IDLE, WALKING, and in VEHICLE.
  Ho et al.~\cite{Ho0SS15} exploited barometer data to infer driving routes.
  And, sudden changes in barometric data was utilized by Wu et al.~\cite{WuPM15} to detect the buildings’ door opening/closing events.

  In a recent work, Quinn~\cite{Quinn19} proposed to use barometers to estimate touch forces 
  in an attempt to provide force sensing functionality in smartphones that lack an integrated force-sensing hardware.
  To this end, Quinn showed that barometers on smartphones with ingress protection can estimate robotic forces on their touchscreen.
  One of the main contributions of our work is to show that barometers on such devices are able to detect even gentle finger taps, 
  and to some extent their positions.
  We further show that barometers on smartphones with ingress protection can reveal information about the phone's earpiece speaker.
  To the best of our knowledge, there is no prior work that studies the feasibility of detecting speaker activities, or detecting sounds in general, 
  using smartphone's barometers.

\newpage

\section{background}
\label{sec:back}

\subsection{Ingress protection}
    Many smartphones are now sealed and nearly air-tight to protect the device's internal components from the ingress of dust and liquid. 
    The degree of protection offered by a device is typically indicated by an IP (Ingress Protection) code from IEC standard 60529.
    An IP rate typically has two digits\cite{Bloch09}.
    The first digit refers to the ingress protection against solids. 
    This scale goes from 0 to 6, where 0 indicates no protection, and 6 indicates dust-tight. 
    The second digit indicates the protection level against water. 
    The digit ranges from 0 to 9, where a higher value correspond with more protection.  
    Most of the mid to high-end smartphones, including the smartphone used in our experiments, are either IP68 or IP67 certified.     
    These ratings imply that they are dust-tight, and protected against immersion in water~\cite{YuXLP19}.

\subsection{Barometer and barometric vents}
    A barometer, also known as air-pressure sensor, is now found in most mid-range to high-end smartphones (e.g. iPhone 6/7/8/X, Samsung note 6/7/8, Google Pixel 2).
    It can act as an altimeter and provide a rough altitude to improve 1) GPS accuracy in vertical direction~\cite{Zhang12};
     2) indoor navigation by revealing floor changes~\cite{MuralidharanKMBA14};
    and 3) accuracy in apps such as fitness apps~\cite{SankaranZGACP14}.

    In devices with ingress protection, barometric vents are added to allow an internal barometer to measure outside atmospheric pressure.    
    This also protects the device’s seals and internal components from the stress that can be placed 
    by the difference in air pressure inside and outside the device.    
    The barometric vents are hydrophobic membrane, and are structured to prevent 
    penetration of dust particle while allowing the flow of air to equalize internal and external barometric pressures.
    The air pressure equalization is, however, slow and occurs over a short period of time.
    This makes it possible to capture sudden changes to the internal pressure using low-rate barometric samples provided by the barometer.

\begin{figure}[htbp]
    \begin{center}
        \includegraphics[scale=0.25]{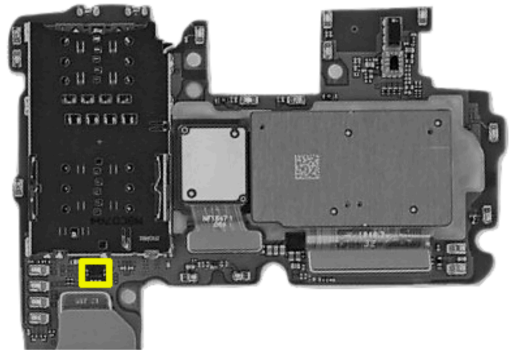}
        \caption{Samsung Galaxy S10 Plus barometer sensor (highlighted).}
        \label{fig:LPS}
    \end{center}
    \end{figure}

    Figure~\ref{fig:LPS} shows the placement of the barometer sensor inside our testing smartphone, Samsung Galaxy S10 Plus.    
    The barometer is an ultra-compact LPS22HH pressure sensor by STMicroelectronics.
    According to the LPS22HH datasheet, the sensor is capable of sampling air pressure at 200 Hz~\cite{LPS}.
    However, the Android Application Programming Interface (API) limits the rate to 25 Hz~\cite{sankaran2014using}.
    In this work, we access the barometer samples at the rate of 25 Hz.

\subsection{Earpiece Speaker}
\begin{figure}[htbp]
    \begin{center}
        \includegraphics[scale=1]{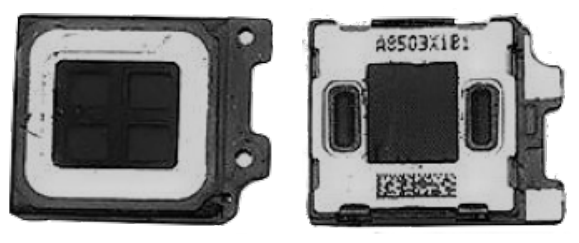}
        \caption{The Samsung Galaxy S10 Plus earpiece speaker with vent holes on both sides to allow air to get in and out.}
        \label{fig:earpiece}
    \end{center}
    \end{figure}

    Figure~\ref{fig:earpiece} shows the Samsung Galaxy S10 Plus earpiece speaker.
    Vent holes on both sides of the speaker allow air to get in and out of the speaker when it plays a sound.  
    Both the earpiece speaker and the barometer are inside the smartphone.
    Therefore, it is natural to ask if the barometer would capture the speaker's activity, as 
    sound is a pressure wave.

\section{Initial observations}
\label{sec:Obs}
    In devices with ingress protection, the internal pressure equalizes with the external pressure over a short period of time.
    This makes it possible to capture changes in the internal pressure using low-rate samples from the device's barometer.
  
\begin{figure}[htbp]
    \begin{center}
        \includegraphics[scale=1.1]{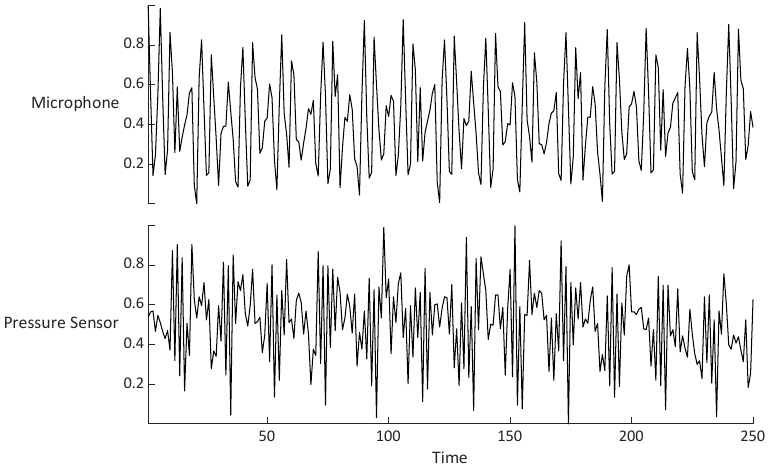}
        \caption{Signals recorded by the device's barometer and microphone when a 12 Hz sinusoid tone was played on the earpiece speaker.}
        \label{fig:sin15}
    \end{center}
    \end{figure}

    Figure~\ref{fig:sin15} shows the barometer samples when a 12 Hz sinusoid tone was played on the device's earpiece speaker. 
    Notice that the barometer sample signal roughly follows the sinusoid signal recorded by the device's microphone.
    This primary observation suggests that it may be possible to detect earpiece speaker activity using the device's internal barometer.

\begin{figure}[htbp]
    \begin{center}
        \includegraphics[scale=0.9]{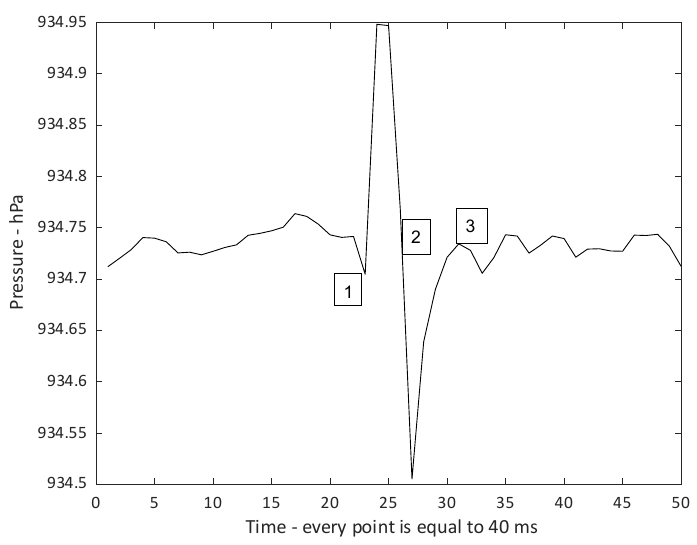}
        \caption{ The impact of a finger tap on the internal pressure captured by the device’s barometer. 
        (1) internal pressure increases as the result of finger tap, (2) a decrease in the pressure after the finger tap, (3) pressure equalization.}
        \label{fig:barSig}
    \end{center}
    \end{figure}

    Figure~\ref{fig:barSig} shows the impact of a touchscreen finger tap on the internal pressure of our testing device.
    A finger tap slightly flexes the screen inwards, causing the internal volume of the device to slightly decrease. 
    Since the device is nearly airtight, the gas inside the device cannot escape immediately through the vents.
    Consequently, the gas inside the device is compressed, causing the internal barometric pressure to increase.
    Once the user removes the finger from the screen, the screen returns to its normal position and causes a short vacuum, 
    this time causing the internal pressure to decrease. 
    This initial observation suggests that low-rate barometer samples may leak information about finger taps.

\pagebreak
\section{Experiments}
\label{sec:Exp}
\subsection{Data collection}
    We developed two custom Android apps for data collection: SpeakerSpy and TouchSpy.
    Using SpeakerSpy, we recorded barometric pressure samples 
    during earpiece speaker activity (i.e., when the given sound is played) and inactivity (i.e., when the speaker is silent).      
    To collect data records, we alternated between ten seconds of speaker activity, 
    and ten seconds of inactivity.
    A two-second ``resting period'' was added between consequent periods of speaker activity an inactivity 
    to allow the internal and external air pressures to equalize.       
    A data record was labeled \emph{active} if the barometer samples were taken 
    during the speaker activity; otherwise, it was labeled \emph{inactive}.
    We collected in total 450 records, half of which were labeled as active and the other half as inactive.
    
    In a second experiment, we used SpeakerSpy to record barometer samples during an external bluetooth speaker activity.
    The objective of this experiment was to check whether the smartphone's barometer can distinguish 
    the case where the external speaker is silent from the case where it plays a sound.
    The external speaker was placed at distance 15cm from the smartphone.
    The sound played by the external speaker was a 12 Hz sinusoid, and the sound level was 73 dBA.
     For this experiment, we collected 270 records.

\begin{figure}[htbp]
    \begin{center}
        \includegraphics[scale=1.1]{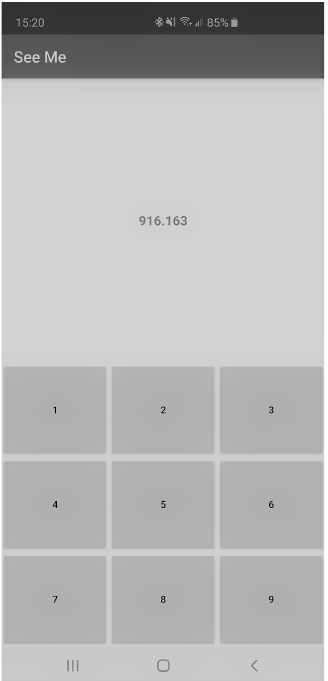}
        \caption{Pin entry keypad (numpad) in the TouchSpy app.}
        \label{fig:key}
    \end{center}
    \end{figure}

    We used the second app, TouchSpy, to read and record barometric pressure samples during a screen finger tap.
    Each record spans an interval of two seconds, starting one second prior to the tap.
    The barometer sampling rate is 25 Hz, thus each record consists of 50 samples.
    %\com{In addition to collecting air pressure samples, the TouchSpy stored the reported sensor accuracy for further processing.}        
    To collect real data using TouchSpy, we employed three participants.
    Each participant was asked to tap (as they normally do) the touchscreen of a stationary phone placed on a desk.
    In one setting, the participants were asked to tap a randomly selected position on the screen.
    In the second setting, the participants were asked to tap a random number on a custom 3-by-3 pin entry keypad shown in Figure~\ref{fig:key}.
    For each setting, we collected 1116 records per participant.
    The records collected in the first setting were used to train and test a classifier for detecting finger taps,
    while the ones collected in the second setting were used to train and test a classifier for inferring the position of a tap.
  
    \textbf{Preprocessing.}  
        We treated each record as a time series, and normalized it to reduce/eliminate the impact of the measurement unit, 
        signal power and gradual changes in atmospheric pressure in the course of data gathering.
        To this end, we used a simple normalization called standardizing, in which each data point $x$ in the time series is replaced by 
        its $z$-score $z=\frac{x-\bar{x}}{S}$, where $\bar{x}$ and $S$ are the mean and standard deviation of the time series, respectively.
        
        For the records corresponding to the external speaker experiment, we used a Savitsky-Golay filter of order 2 and frame size 5 
        to reduce the effect of noise.
        This additional preprocessing improved our detection accuracy by 10\% (from 82\% to 92\%).

    \textbf{Classificaiton and performance evaluation.}
        We used a support-vector machine (SVM\footnote{SVM is a popular classification method, which gets a set of $n$-dimensional points, 
        and constructs a hyperplane or a set of hyperplanes to separate them.})  as our classification method.
        To evaluate the performance of SVM, we applied multiple 5-fold cross validations. 
        In a 5-fold cross validation, the records are randomly split into five groups of equal size, 
        where each group has equal number of records from each class.
        Then, for every number~$i$, $1\leq i \leq 5$,
        \begin{enumerate}
          \item the group $i$ is marked as the \emph{test set};
          \item the remaining four groups are marked as the \emph{training set};
          \item the SVM classifier is trained using the training set;
          \item the accuracy of the trained classifier is measured using the test set.        
        \end{enumerate}
        At the end of the above iterative process, the average of the five accuracy scores (obtained in Step 4) is returned as the accuracy result
        of the cross validation. 
        In this work, every reported accuracy result is the average of the results of ten 5-fold cross validations.

\subsection{Accuracy results}
%\subsection{Assumptions and limitations}
    \textbf{Earpiece speaker activity detection.}
        Barometer-based speaker activity detection (B-SAD) is a binary classification problem:
        given a sound and the barometer samples, infer whether the smartphone's speaker is playing the sound or is silent.
        Note that a malicious app with B-SAD capability can steal user's sensitive information. For example, it can set the sound to, say, 
        the device's default ringtone to detect when a phone call is received.
        
        To evaluate the performance of SVM in solving B-SAD, we used two types of sounds.
        The first one was a simple sinusoid tone of 12 Hz (nearly half the barometer's sampling rate of 25 Hz). 
        For the second sound we used a popular Android ring tone. 
        In our experiment, the SVM classifier achieved a detection accuracy of 96\% for the 12 Hz sinusoid tone, and 95\% 
        for the default Samsung ringtone\footnote{Over-the-horizon ringtone, extracted from Samsung Galaxy S10 Plus.}.
        For the case of external speaker activity, SVM achieved an accuracy of 92\%.
        We remark that these are the first feasibility results showing that  
        barometer samples can leak information about both internal and external speaker activity. 
        
        Our testing smartphone is IP68 rated (i.e., a good level of ingress protection), which makes
        it possible to detect speaker's activity using low-rate barometer samples.        
        We remark that barometer samples may leak information about the speaker's activity in non-IP rated smartphones, too.
        To show this, we repeated the above experiment with Samsung Galaxy S6 Edge, which is a non-IP rated smartphone.
        In this experiment, SVM achieved a detection accuracy of 67\%.
        This accuracy is non-negligible, although it is significantly lower than the 95\% accuracy we achieved in the IP68-rated smartphone.

    \textbf{Detecting taps.}  
        Quinn~\cite{Quinn19} recently showed that, in devices with ingress protection, 
        forces on touchscreen can be detected using the device's internal barometer.
        To  this end, a robotic arm was programmed to perform five second touch contacts with three different force levels (1, 3, and 5N).
        In our work,  however, we explore the possibility of detecting finger taps (rather than robotic arm pressures) and their positions. 
        %In this work, we extend Quinn's result by replacing robotic touch contacts with user's normal finger touch contacts. 
        
        Unlike the 5-second robotic arm touch contacts in~\cite{Quinn19}, finger taps have a short touchscreen contact time.
        In our experiment, the touchscreen contact time of a finger tap ranged from 24ms to 183ms, with the average of 85ms.
        As stated in~\cite{Quinn19} if a user’s input force changes too quickly, it may not be captured by the barometer.
        In addition, based on the study in~\cite{TaherAHV14}, finger taps typically produce less force than those applied in Quinn's experiment.        
        Therefore it is not clear if finger taps are detectable at all. 
        In our experiment, however, the SVM classifier achieved 100\% accuracy in detecting finger taps.        
        %Finding the position of a tap is, however, much harder than detecting the tap.
        Next we show that barometer samples can reveal information about the position of taps, too.

    \textbf{Detecting the position of a tap.}     
        As mentioned earlier, we used a custom numpad to collect finger tap records.        
        This numpad (Figure~\ref{fig:key}) uses exclusively the lower half of the screen, 
        as this is the default location of touch keyboards and numpads on most smartphones. 
        The numpad divides the lower half of the screen into nine different regions/keys. 
        Each region/key has different degrees of freedom.
        The difference in degrees of freedom leads to different amounts of screen flex and recovery as the result of a tap. 
        This makes it possible to profile each region/key, and guess the number which was tapped.
         
\begin{figure}[htbp]
    \begin{center}
        \includegraphics[scale=0.6]{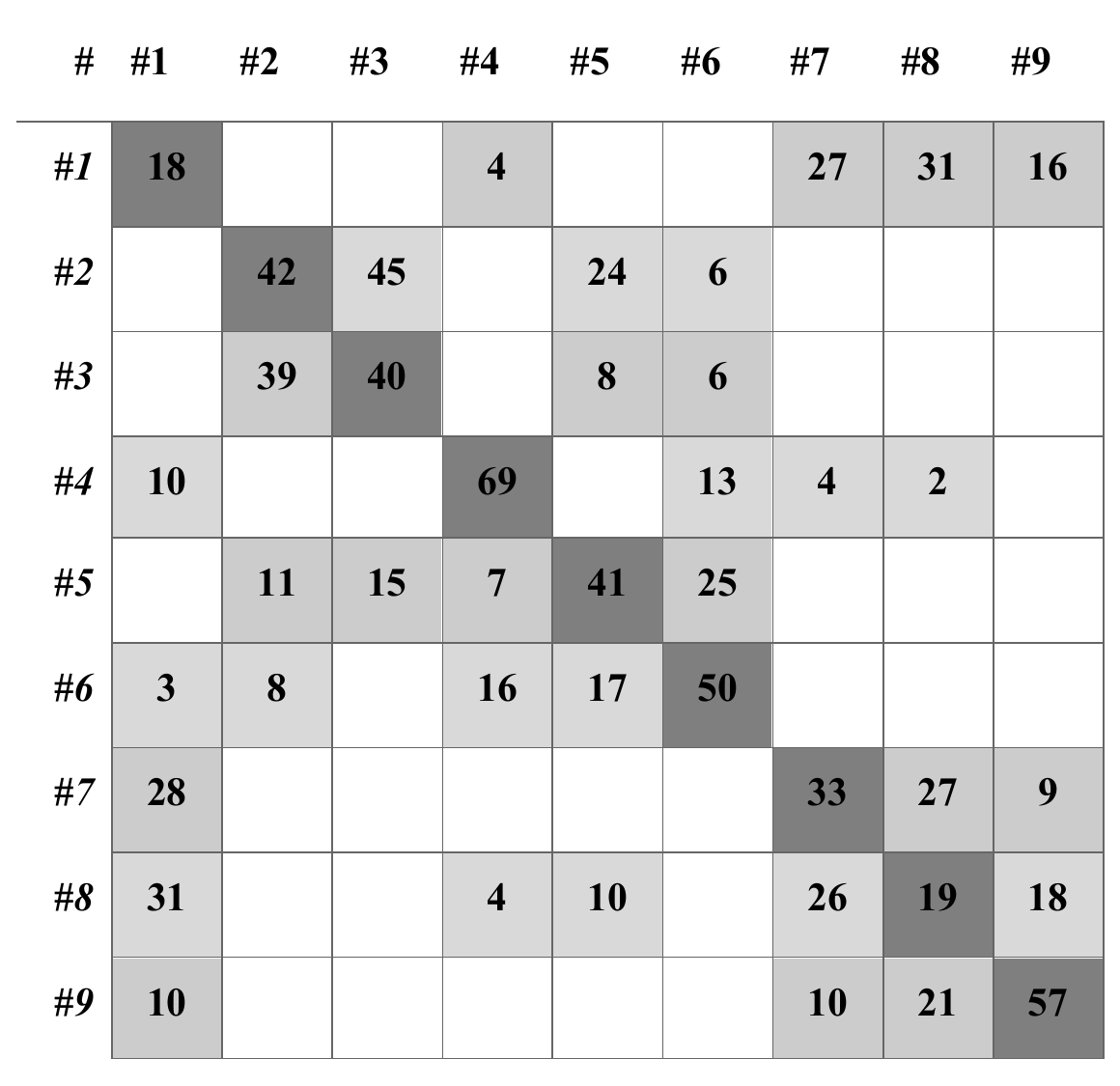}
        \caption{The number in row $i$ and column $j$ of the table is the probability that the classifier returns $i$ 
          when the actual number tapped was $j$. }
        \label{fig:t1}
    \end{center}
    \end{figure}

        Our accuracy results are summarized in Figure~\ref{fig:t1}.
        The number in row $i$ and column $j$ of the table is the probability the the classifier outputs $i$ when the actual number tapped was $j$.
        For instance, when user presses 5 our SVM classifier outputs 2, 3, 5, 6 and 8 with probabilities 24\%, 8\%, 41\%, 17\% and
        10\%, respectively.
        As shown in Figure~\ref{fig:t1}, probabilities of correct identification (i.e., numbers on the main diagonal of the table) 
        range from $18\%$ to $69\%$ with the average of $41\%$.
        One may compare this to the probability of correct identification in random guess, which is about $11\%$.

  %Clearly, this is not a perfect detection, but still can be used in an attack scenario to narrow down the position of screen touch. 

\section{An attack scenario}
    Traditional side-channel attacks such as~\cite{van1985electromagnetic} require the adversary to have physical access 
    to the device, or be in the close vicinity of the device. 
    The trend has, however, shifted for smartphones. 
    Using the results presented in this work, an adversary can execute a barometer-based side-channel attack entirely remotely. 
    To do so, the adversary needs to take the following steps to execute a successful touchscreen tap inference or B-SAD on a phone.

    \begin{enumerate}
	\item{\textbf{Embed a malicious code into an app }}. 
	  An adversary has at least two options to access the barometer's API. 
	  The first option is to build a new app (e.g., a weather forecast app) from scratch, 
	  and embed the malicious code (which gathers barometer samples, and performs inference) inside it. 
	  The second option is to inject the malicious code into a popular and well-known app such as a popular game. 
	  There are pros and cons to each method. 
	  Making an app from scratch gives the attacker the ability to publish the malicious app silently via, say, the Google Play Store. 
	  Also, it gives the attacker the chance to request some permissions (like Internet access) that may initially look legitimate. 
	  Such accesses can be beneficial. 
	  For example, the Internet access can be used to transfer the inferred user's information to the attacker. 
	  By injecting the malicious code into a well-known and popular app, on the other hand, 
	  the attacker may hit a larger population. 
	  This option, however, does not allow the attacker to put the modified app on a trusted source like Google Play Store. 
	  This makes it harder for the app to find its way to the users’ phones.
    	\item{\textbf{Perform data gathering and training}}. 
	  Data gathering does not require any permissions from the user, but needs the user to interact with the malicious app. 
	  With an invisible overlay layout on top of the malicious app, the app can gather the necessary data for training 
	  a model for the user's typing habits (the exact force of user's finger taps) and phone screen size.
	  In our tests, about 50 samples per region (key) was the minimum for our classification to be viable.
	  As for B-SAD, embedded sound queues in the malicious app are enough to train the model. 
	  In our tests, to perform an accurate B-SAD, a minimum of 25 samples of each state (active/silence) is necessary.
	  An adversary can use the gathered data to train an SVM classifier either locally at the victims' smartphones, 
	  or remotely at their own end.
	\item{\textbf{Perform inference and extraction}}. 
	  At this point, the malicious app continuously logs the barometer samples in the background. 
	  At the same time, it feeds the collected samples to its SVM classifier to detect taps and/or earpiece speaker activity. 
	  The results of these detections can be stored locally, or sent to the adversary over the Internet. 
\end{enumerate}

\subsection{Limitations} 
    Our results confirm the possibility of detecting earpiece speaker activity and touchscreen taps using low-rate barometric air pressure samples. 
    These results rely on relative pressure rather than absolute pressure values of the barometer; thus they are applicable 
    to different locations/altitude levels, and weather conditions.      
    However, there are certain limitations in detecting speaker activity and finger taps using low-rate barometer samples.
    First, the device must have some level of ingress protection.
    It is because, otherwise, changes in internal pressure (due to, for example, finger taps) would quickly equalize the external atmosphere.
    
    The second major limitation is related to the low sampling rate of the barometer sensor. 
    The STMicro LPS22HH sensor datasheet shows that the sensor is capable of sampling at 200 Hz. 
    However, the Android sensor API limits the barometer sample rate to 
    25 Hz\footnote{The sampling rate of 25 Hz is provided by the SensorManager.SENSOR DELAY FASTEST in Android SDK.}~\cite{sankaran2014using}. 
    In our experiments, the low sampling rate of 25 Hz was not deterrent in detecting earpiece speaker activity and detecting finger touch contacts. 
    There are, however, certain tasks that are impossible to perform at low sampling rates.    
    For instance, by the Nyquist–Shannon sampling theorem, it is impossible to reconstruct the speaker's sound signal fully 
    if the signal has frequencies higher than half the sampling rate of the barometer.
    
    The low sampling rate of barometer can be preventive in carrying certain tasks related to screen touch detection as well.  
    For example, in our experiment, we need to get at least three sample points in our signal to register a finger tap, 
    and then wait for at least two sampling cycles for the internal and external air pressures to equalize. 
    At the sampling rate of 25 Hz, these five sampling cycles translate to a total of 200 milliseconds, 
    which would limit the continuous detection of finger taps to five taps per second.    
    
    Finally, the flow rate of the barometric vent can also become a limiting factor.
    As stated in~\cite{Quinn19}, if a user's input force changes either too quickly or too slowly, it may not be captured by the barometer.

\section{Security Implications}
\label{sec:Sec}
  Reading barometric pressure low-rate samples is considered harmless.
  Therefore, modern mobile platforms like Android allow apps to access the device's barometer sensor without user's permission or notification.
  Consequently, a malicious app can readily bypass the user's permission and attention.    
  In addition, as shown in~\cite{SimonA13}, sensor-based side-channel attacks can bypass strong separation mechanisms like Samsung KNOX, 
  which try to provide a secure environment to protect corporate data on smartphones.
  
  Finally, due to barometer's low power usage,
  accessing the barometric pressure samples may not be detected as, for example, a power virus. 
  We ran a background service that continuously reads barometric pressure samples for over 40 hours.
  This service was not detected as a suspicious activity by McAfee, Avast, AVG, Bitdefender and Norton antivirus.

\section{Defenses} 
\label{sec:Def}
  A first line of defense is to use the device's permission system to ask users for permission to access the device's barometer. 
  This gives users the chance to deny permission to apps whose motivation behind requesting the access is not clear.
  For instance, users can deny barometer access to a ``flashlight app'' as such access is clearly outside the scope of the app's functionality. 

  Another defense is to use third party apps such as AppGuard~\cite{BackesGHMS13} that enforce security policies.
  These apps can pause sensor readings (or even kill) background processes/services 
  when an app with sensitive input (e.g. a banking app) is running.
  In addition, malware-analysis apps such as~\cite{FeltCHSW11} and~\cite{GiblerCEC12} can be extended and used to check for malicious sensor accesses.
  
  Finally, sensitive apps can implement their own custom security solution.
  For example, a PIN pad can rearrange its buttons prior to every sensitive input~\cite{OwusuHDPZ12}.

\section{Conclusion and future work}
\label{sec:Con}
    Speaker and touchscreen input contain sensitive information.
    Because of this, touchscreen input is only given to the foreground app,
    and microphone's access (which reveals the speaker activity) requires user permission.
    On the other hand, any app  can access the barometer samples without any permission or notification. 
    This is alarming as our results indicate that low-rate barometric pressure samples reveal information about earpiece speaker activity.
    In addition, the pressure samples leak information about finger taps, and their positions.
%    This is alarming because any app (even those running in the background) 
%    can access the barometer's samples without any permission or notification.   
    
    Our work is a first feasibility study, and there are many ways to extend and improve our results.    
    First, we only used SVM as our classifying method. 
    More advanced classifiers and techniques can improve our reported accuracy results.  
    In addition, barometric pressure samples may be combined with data samples from other sensors such as ambient-light or motion sensors 
    to improve accuracy. 
    
    In detecting finger taps and their positions, we used a small number of participants to collect data. 
    Further study is needed to show if similar results are observed with more number of participants with, say, various tapping habits.
    Also, we used a single type of smartphone in our experiments.
    We, however, expect similar results to be observed in many other types of smartphones with ingress protection.    
    
    With regards to earpiece speaker activity, it is interesting to see how many different classes of activities we can distinguish using 
    low-rate barometric pressure samples.
    Our initial observations, not reported to this work, indicates that it is possible to distinguish different sounds.
    
    Finally, it is interesting to see how far we can go in detecting external sounds using low-rate barometer samples.

\newpage
\bibliographystyle{IEEEtran}
\bibliography{references}   

\end{document}